\documentclass[prl,twocolumn,superscriptaddress,floatfix,nofootinbib]{revtex4-2}

\usepackage{graphicx}
\usepackage{dcolumn}
\usepackage{tensor}
\usepackage{bm}
\usepackage{float}
\usepackage{color}
\usepackage[usenames,dvipsnames]{xcolor}
\usepackage{listings}
\usepackage{fancyvrb}
\usepackage{slashed}
\usepackage{amsmath}
\usepackage{amssymb}
\usepackage{amsfonts}
\usepackage{enumitem}
\usepackage{listings}
\usepackage{mathtools}
\usepackage{slashed}
\usepackage{verbatim}
\usepackage{hyperref}
\usepackage{braket}
\usepackage[margin=1.0in]{geometry}
\usepackage{slashed}

\usepackage{accents}
\makeatletter
\DeclareRobustCommand{\cev}[1]{%
  \mathpalette\do@cev{#1}%
}
\newcommand{\do@cev}[2]{%
  \fix@cev{#1}{+}%
  \reflectbox{$\m@th#1\vec{\reflectbox{$\fix@cev{#1}{-}\m@th#1#2\fix@cev{#1}{+}$}}$}%
  \fix@cev{#1}{-}%
}
\newcommand{\fix@cev}[2]{%
  \ifx#1\displaystyle
    \mkern#23mu
  \else
    \ifx#1\textstyle
      \mkern#23mu
    \else
      \ifx#1\scriptstyle
        \mkern#22mu
      \else
        \mkern#22mu
      \fi
    \fi
  \fi
}

\makeatother

\def\ee{\end{eqnarray}}

\def\nn{\nonumber}

\newcommand{\be}{\begin{eqnarray}}
\newcommand{\en}{\end{eqnarray}}
\newcommand{\bea}[1]{\left(\begin{array}{#1}}
\newcommand{\ena}{\end{array}\right)}
\newcommand{\ba}{\begin{eqnarray}}
\newcommand{\ea}{\end{eqnarray}}

\newcommand{\CO}{\mathcal{O}}

\newcommand\lrnab{\raise .8ex\hbox{$^\leftrightarrow$} \hspace{-8.8pt}
\nabla}
\newcommand\lnab{\raise .8ex\hbox{$^\leftarrow$} \hspace{-9.8pt}
\nabla}
\newcommand\rnab{\raise .8ex\hbox{$^\rightarrow$} \hspace{-9.8pt}
\nabla}

\begin{document}


\title{Nuclear-Level Effective Theory of $\mu \rightarrow e$ Conversion}

\author{Evan Rule}
\affiliation{Department of Physics, University of California, Berkeley, CA 94720, USA}
\author{W. C. Haxton}
\affiliation{Department of Physics, University of California, Berkeley, CA 94720, USA}
\affiliation{Lawrence Berkeley National Laboratory, Berkeley, CA 94720, USA}
\author{Kenneth McElvain}
\affiliation{Department of Physics, University of California, Berkeley, CA 94720, USA}

\date{\today}

\begin{abstract}
The Mu2e and COMET $\mu \rightarrow e$ conversion experiments are expected to significantly advance limits on new sources of
charged lepton flavor violation (CLFV).   Almost all theoretical work in the field has focused on just two operators. However, general symmetry arguments
lead to a $\mu \rightarrow e$ conversion rate with six response functions, each of which, in principle, is observable by varying nuclear properties of
targets.  We construct a nucleon-level nonrelativistic effective theory (NRET) to clarify the microscopic origin of
these response functions and to relate rate measurements in different targets.  This exercise identifies three operators and their small parameters 
that control the NRET operator expansion.  We note inconsistencies in past treatments of these parameters.
The NRET is technically challenging, involving 16 operators, several distorted electron
partial waves, bound muon upper and lower components, and an exclusive nuclear matrix element.  We introduce a trick for treating the
electron Coulomb effects accurately, which enables us to include all of these effects while producing transition densities whose one-body
matrix elements can be evaluated analytically, greatly simplifying the nuclear physics.
We derive bounds on operator coefficients
from existing and anticipated $\mu \rightarrow e$ conversion experiments.  We discuss how similar NRET formulations have impacted dark matter phenomenology, noting that
the tools this community has developed could be adapted for CLFV studies.
\end{abstract}


\maketitle
$\mu \rightarrow e$ conversion and other charged lepton flavor violation (CLFV) processes have long been recognized as sensitive tests of new physics beyond the standard model \cite{Snowmass1,Barbieri,Cirigliano,Calibbi}.  The Mu2e \cite{Mu2Eover,Abusalma} experiment at Fermilab and the COMET \cite{COMETI,COMETII} experiment at J-PARC aim to improve existing limits on the ratio of rates
\begin{equation}
R_{\mu e} =\frac{\omega\left[\mu^-+(A,Z)\rightarrow e^-+(A,Z)\right]}{\omega\left[\mu^-+(A,Z)\rightarrow \nu_{\mu}+(A,Z-1)\right]}
\end{equation}
by four orders of magnitude, probing new physics to scales of $\sim 10^4$ TeV.
Both experiments will use the relatively light target $^{27}$Al and focus on captures that leave the nucleus in its ground state.  The elastic process
maximizes the energy of the outgoing electron, $E_e \sim m_\mu-E_\mu^\mathrm{bind}$ neglecting nuclear recoil, where $E_\mu^\mathrm{bind}$ is
the (positive) muon binding energy.  This minimizes the standard-model background from the three-body decay  $\mu^- \rightarrow e^- + \nu_\mu + \bar{\nu}_e$, as few electrons are produced near the endpoint.  

Consequently, the CLFV operators contributing to $\mu \rightarrow e$ conversion
are constrained by the nearly exact parity and CP of the nuclear ground state: these symmetries eliminate some CLFV operators entirely and
restrict the contributing multipoles of those that do survive.  Our aim is to determine the constraints that can be
imposed on the coefficients of the operators of a general nucleon-level effective theory (ET) of CLFV, thereby identifying what can and cannot be learned
from elastic $\mu \rightarrow e$ conversion.

The form of the $\mu \rightarrow e$ rate can be deduced from general considerations, including Galilean/rotational invariance,
the good parity (P) and CP of the nuclear ground state, and the symmetry transformation properties of the available charges and currents,
and their gradients and curls.  The detailed derivation is included in an accompanying technical paper \cite{HRMR}.   The resulting rate
consists of six response functions (and two interference terms), each of which is, in principle, observable by making measurements
in multiple nuclear targets selected for their ground-state properties, allowing one to emphasize or suppress certain interactions.  


Rates in different nuclei can be related using a nucleon-level interaction, the general form of which is determined by nonrelativistic effective theory (NRET).   One constructs a general CLFV interaction from the available operators --- including, in this case, the charge and spin operators for the leptons
and nucleons, the nucleon relative velocity operator $\vec{v}_N$ (conjugate to the internucleon coordinate $\vec{r}$), and the muon velocity operator $\vec{v}_{\mu}$ (conjugate to the coordinate describing the bound muon relative to the nuclear center-of-mass).   The construction
follows that performed for dark matter (DM) phenomenology \cite{Liam1,Liam2,Snowmass2}. When this interaction is
embedded in the nucleus, an additional operator enters due to nuclear compositeness, $\vec{q} \cdot \vec{r} \sim 1$.
As in DM direct detection, the three-momentum transfer $q$ is comparable to the nuclear size, implying significant angular momentum transfer.
Thus the nuclear physics is complex, requiring a multipole expansion.

NRET predictions of low-energy observables will be equivalent to those of higher-energy EFTs (if both theories are expanded to the same order),
but as fewer degrees of freedom are relevant at low energies, the NRET operator basis will be leaner and more efficient.  The information
extracted from experiment can then be ported upward, constraining higher level theories.  A conceptual depiction of the
process is captured in Fig. 4 of a DM Snowmass report \cite{Snowmass2}, where considerable work has been invested in matching theories,
most of which can be readily adapted to $\mu \rightarrow e$ conversion.  

In NRET, one identifies a theory's small parameters and organizes the operator expansion accordingly.   We find, associated with $\vec{q} \cdot \vec{r}$, $\vec{v}_N$, and $\vec{v}_\mu$, the expansion parameters 
\begin{equation*}
y=\left(  { q b \over 2} \right)^2=0.24 > | \langle {\vec{v}_N  \over 2} \rangle | =0.11 >  | \langle {\vec{v}_\mu  \over 2} \rangle | =0.026 
\end{equation*}
where the numerical values are those of Al.  Here $b$ is the oscillator parameter representing the nuclear size.  The hierarchy of operators and the expansion parameters they generate should guide
all treatments of $\mu \rightarrow e$ conversion, but this typically has not been the case.  Table 1 of \cite{HRMR} provides a summary of the literature.
Rate calculations have been done repeatedly for just two nuclear operators, $1_L 1_N$ and $\vec{\sigma}_L \cdot \vec{\sigma}_N$.  The
velocity operator $\vec{v}_\mu$, which generates the muon's lower component, has been explored frequently but always with a restriction on the included electron partial waves of $|\kappa|=1$, where $\kappa$ is the Dirac quantum number,
thereby limiting the expansion in $y$ so that only the leading nuclear multipole is retained.  This is done because otherwise the
algebra is complicated, but the restriction --- parametrically and numerically --- typically generates an error larger than the $\vec{v}_\mu$ correction being made \cite{HRMR}.  Here, we describe a formalism that treats $\vec{v}_\mu$ accurately and elegantly, including all multipoles.   We
find that $\vec{v}_\mu$ is a relatively uninteresting operator, playing no symmetry role and contributing only a 5\% numerical
correction to Al nuclear form factors.

In contrast, none of papers in Table 1 of \cite{HRMR} treats $\vec{v}_N$.  This operator, parametrically larger than $\vec{v}_{\mu}$, 
represents the $A-1$ internal (Jacobi) internucleon velocities in the bound state.  (The $A$th Jacobi velocity is $\vec{v}_\mu$.)   Without the inclusion of $\vec{v}_N$, the NRET rate is not consistent with the
general form derived from symmetry arguments: Potentially important physics has been omitted.  
Further, this operator produces a novel form of coherence that operates only in certain nuclei like Al and Cu, where the Fermi sea
fills only one of two spin-orbit subshells,  $(\ell s)j=\ell\pm {1 \over 2}$.   While one would expect $\vec{v}_N$-dependent operators to be
subleading, of  $o\left({q \over m_N}\right)  \sim {1 \over 10}$, this coherence can elevate certain velocity-dependent operators to o(1),
where they can dominate rates even if operators like $1_N$ and $\vec{\sigma}_N$ are also present.

{\it Leptonic Interactions:}
$\mu \rightarrow e$ conversion proceeds through the capture of a 1s muon bound by the nuclear Coulomb field and the emission of a highly relativistic electron.  It is desirable, especially
in heavier targets,  to obtain the  lepton wave functions from the Dirac equation, using an extended nuclear charge distribution \cite{Czarnecki, Kitano, Crivellin, Bartolotta, Cirigliano:2022ekw}.  The transition 
density then depends on a convolution of a nuclear density (which depends on the operator being studied), the muon wave functions, and various electron distorted waves that, in the 
absence of the Coulomb interaction, would correspond to the spherical wave expansion of $e^{i \vec{q} \cdot \vec{r}}$.

The muon quickly cascades to the  $1s$ orbital as it comes to rest in a target.  The Dirac solution is related to the muon's velocity operator through the approximate relationship \cite{HRMR} 
 \be
 \psi_{\kappa=-1} =  \left[ \begin{array}{c} i g(r)  \xi_s \\ f(r)  { \vec{\sigma} \cdot \hat{r}  }  \xi_s\end{array}  \right]{  1 \over \sqrt{4 \pi}} \sim  \left[ \begin{array}{c}  \xi_s \\  { \vec{\sigma} \cdot \vec{v}_\mu \over 2 }  \xi_s\end{array}  \right]{  i g(r) \over \sqrt{4 \pi}} ~~~
 \label{eq:muonspinor}
 \ee
where $\xi_s$ is the Pauli spinor. The  normalization is
\begin{equation*} 
\int_0^\infty dr\; r^2 (g^2_\kappa(r) + f^2_\kappa(r)) =1.
\end{equation*}
A linear expansion in $\vec{v}_\mu$ means retaining $f(r)$.

\begin{figure}   
\centering
\includegraphics[scale=0.42]{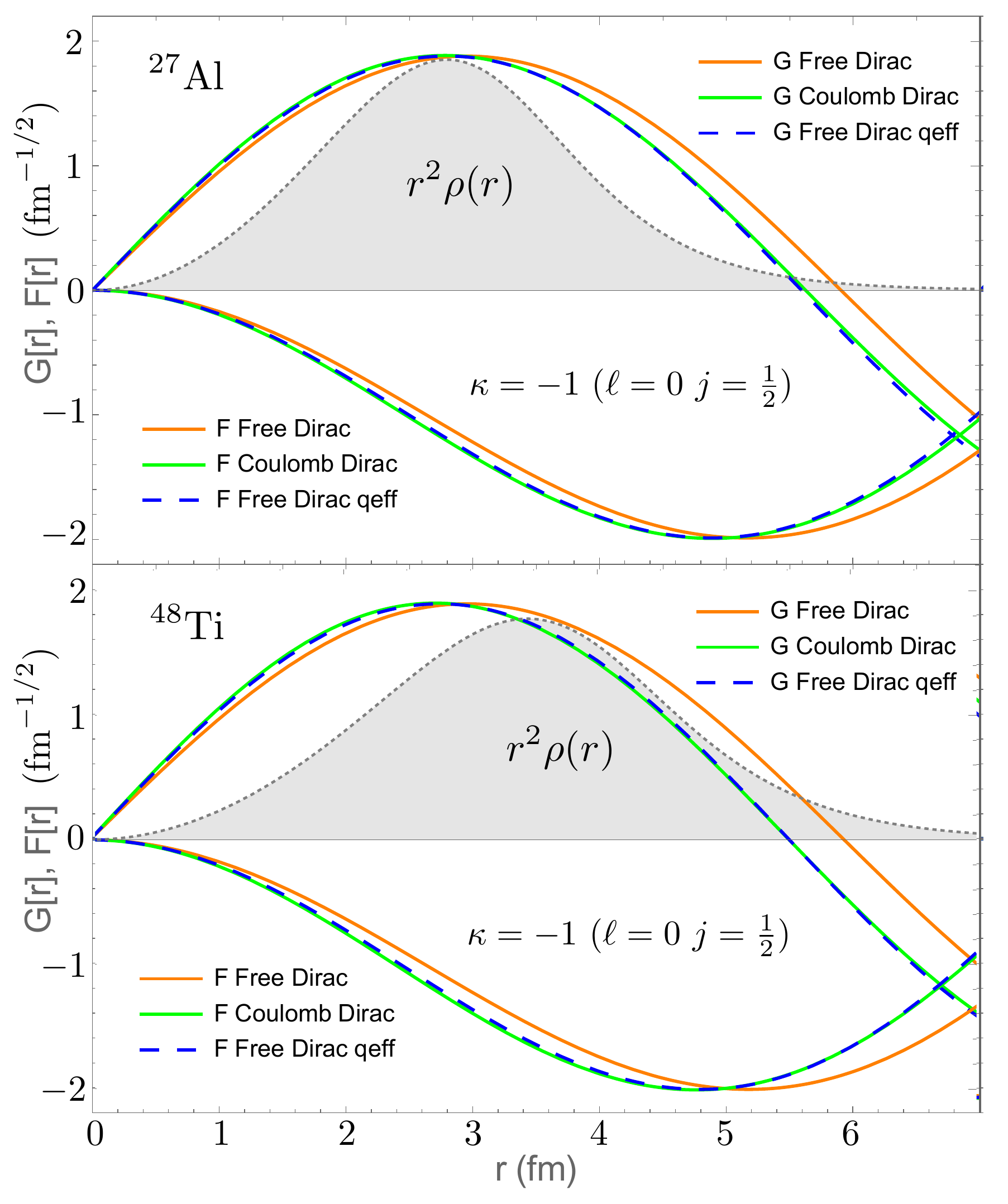}
\caption{Free and Coulomb Dirac solutions $G(r)= r g(r)$ and $F(r)=r f(r)$ are compared to the $q_\mathrm{eff}$-shifted free Dirac solution
for the relativistic electron. The nuclear charge distribution (shaded) used in the Coulomb calculation is also shown (arbitrary normalization).}
\label{fig:Ae1}
\end{figure}

The electron produced in $\mu \rightarrow e$ conversion is ultra-relativistic.  Deviations from a Dirac plane wave arise from the nuclear Coulomb attraction, which modestly enhances the wave function amplitude 
near the nucleus and shortens the wavelength. We account for these effects by a procedure familiar from electron scattering studies \cite{Dirac12,Dirac13,Dirac15}, replacement of $\vec{q}$ in the Dirac plane wave by a shifted momentum
$\vec{q}_\mathrm{eff}$, yielding
\be
\displaystyle{ U(q,s)~e^{i \vec{q} \cdot \vec{x}}  \rightarrow  \sqrt{E_e \over 2 m_e} \left( \begin{array}{c} \xi_x \\ \vec{\sigma} \cdot  \hat{q} \xi_s \end{array} \right){q_\mathrm{eff} \over q}  e^{i \vec{q}_\mathrm{eff} \cdot \vec{x}}}.
 \label{eq:electronspinor}
 \ee
 The local momentum $q_\mathrm{eff}$ is obtained from a constant potential whose depth is equated to the average of the Coulomb potential over the nuclear charge distribution.  We find  $q_\mathrm{eff}=110.81$ MeV and $112.43$ MeV for $^{27}$Al and $^{48}$Ti, respectively. The effective momentum approximation, the Dirac plane-wave solution, and the Dirac Coulomb solution for $^{27}$Al and $^{48}$Ti are shown in Fig. \ref{fig:Ae1}.   The $q_\mathrm{eff}$ approximation nicely
 accounts for the distortion.   The results shown correspond to the $\kappa=-1$ state, but this approximation works equally well for other Dirac partial waves and remains accurate
 for high-$Z$ targets of interest, such as  W \cite{HRMR}.
 
 
{\it The NRET construction:}  Eqs. (\ref{eq:muonspinor}) and (\ref{eq:electronspinor}) allow one to construct a nucleon-level NRET for $\mu \rightarrow e$ conversion that 
operates between Pauli spinors. Available operators include the lepton and nucleon
identities $1_L$ and $1_N$ and five dimensionless Hermitian three-vectors
 \ba
 && i \hat{q} ={i \vec{q} \over |\vec{q}\,|}, \ \ \ \ \vec{\sigma}_L, \ \ \ \ \vec{\sigma}_N,  \ \ \ \ \vec{v}_N,  \ \ \ \ \vec{v}_\mu
 \ea
The nuclear velocity operator has the symmetrized form $\vec{v}_N=(\vec{p}_i+\vec{p}_f)/(2m_N)$.

Omission of the velocity operators will produce a rate appropriate for a point nucleus, but
not generate all projections of the vector current allowed by symmetry \cite{HRMR}. The addition of $\vec{v}_N$ completes the NRET, and interesting new physics arises when $\vec{v}_N$-dependent operators are  embedded in the nucleus. In contrast, $\vec{v}_\mu$'s only role is to generate nuclear form factor corrections suppressed by the ratio of average values of the muon's lower and upper components $\langle f \rangle/\langle g \rangle$.
We will omit this operator presently.

The NRET includes  $\vec{v}_N$  only linearly, as certain ambiguities arise in bound-state applications beyond first order \cite{Serot}. The nucleon operators can be combined as 
$\vec{v}_N \cdot \vec{\sigma}_N$ and $\vec{v}_N \times \vec{\sigma}_N$, but not as the rank-two tensor
$\left[ \vec{v}_N \otimes \vec{\sigma}_N \right]_2$, 
which would not triangulate between spin-$\textstyle{1 \over 2}$ nucleon states. As the interaction occurs at a fixed $q \sim m_\mu$, any propagator effects can be absorbed into the operator coefficients.  

We identify a total of 16 independent operators
\allowdisplaybreaks
\be
\CO_1 &=& 1_L ~1_N ~~~~~~~~~~~~~~~~ \,\CO_9 = \vec{\sigma}_L \cdot \left( i \hat{q} \times \vec{\sigma}_N \right) \nn \\
\CO^\prime_2 &=& 1_L ~i \hat{q} \cdot \vec{v}_N ~~~~~~~~~~~\,\CO_{10} = 1_L~ i \hat{q} \cdot \vec{\sigma}_N \nn\\
 \CO_3 &=& 1_L~  i \hat{q} \cdot  \left[ \vec{v}_N \times \vec{\sigma}_N \right] ~\,\CO_{11} = i \hat{q} \cdot \vec{\sigma}_L ~ 1_N  \nn \\
\CO_4 &=& \vec{\sigma}_L \cdot \vec{\sigma}_N ~~~~~~~~~~~~~~\,\CO_{12} = \vec{\sigma}_L \cdot \left[ \vec{v}_N \times \vec{\sigma}_N \right] \nn\\ 
\CO_5 &=&  \vec{\sigma}_L \cdot \left( i \hat{q} \times \vec{v}_N \right) ~~~~~\,\CO^\prime_{13} = \vec{\sigma}_L \cdot  \left( i \hat{q} \times \left[ \vec{v}_N \times \vec{\sigma}_N \right] \right)  \nn \\
 \CO_6&=&  i \hat{q} \cdot \vec{\sigma}_L ~ i \hat {q} \cdot \vec{\sigma}_N ~~~~~~\CO_{14} = i  \hat{q} \cdot \vec{\sigma}_L ~ \vec{v}_N \cdot \vec{\sigma}_N  \nn \\
\CO_7 &=& 1_L~ \vec{v}_N \cdot \vec{\sigma}_N ~~~~~~~~~\,\CO_{15} = i \hat{q} \cdot \vec{\sigma}_L~ i \hat{q} \cdot \left[ \vec{v}_N \times \vec{\sigma}_N \right] \nn\\
\CO_8 &=& \vec{\sigma}_L\cdot \vec{v}_N ~~~~~~~~~~~~~~\,\CO^\prime_{16} = i \hat{q} \cdot \vec{\sigma}_L ~i \hat{q} \cdot  \vec{v}_N 
\label{eq:ops}
\ee
With the inclusion of isospin to allow for distinct proton and neutron couplings, the ET Lagrangian takes the form 
\begin{equation}
\mathcal{L}=\sqrt{2}G_F\sum_{\tau=0,1}\sum_{i=1}^{16}\tilde{c}_i^{\tau}\mathcal{O}_it^{\tau}
\label{eq:et_interaction}
\end{equation}
where $t^0=1$, $t^1=\tau^3$. The coefficients $\tilde{c}_i$ are the low-energy constants (LECs) that contain the CLFV physics. The tildes denote  that $\tilde{c}$ and  $\tilde{R}$ (see below) are dimensionless and normalized to the scale $\sqrt{2}G_F$.

{\it The Nuclear Embedding:}  The embedding of an NRET in a nucleus can have interesting consequences in exclusive processes, enhancing sensitivity to some operators,
reducing through selection rules sensitivity to others.  The embedding also alters the interpretation of the LECs:  various corrections associated with more complicated currents, operator 
mixing that arises from the incomplete nuclear Hilbert space, and other
many-body effects can shift LEC values. The NRET's complete operator basis can accommodate mixing effects that redistribute strength among the operators.
The LECs remain a very useful way to represent the CLFV physics extracted from experiment, including correlating results from different targets, but the extracted LECs require some interpretation \cite{HRMR}.
  
One proceeds by generating one-body charge and current operators from Eq. (\ref{eq:et_interaction}) and performing
a standard multipole analysis \cite{Walecka}.  While this generates 11 nuclear response functions $W_{\mathcal{O}}^{\tau\tau'}$, five are eliminated by parity and CP
constraints on elastic scattering \cite{Liam2,HRMR}.  The $\mu\rightarrow e$ decay rate is
\allowdisplaybreaks
\begin{eqnarray}
\label{eq:rate}
\omega =&& {G_F^2 \over \pi} ~ {q_\mathrm{eff}^2 \over 1+{q \over M_T}}  ~ |{\phi}^{Z_\mathrm{eff}}_{1s}|^2  ~ 
\sum_{ \tau=0,1} \sum_{\tau^\prime = 0,1} \nonumber \\
&&\times \Bigg\{  ~\left[ \tilde{R}_{M}^{\tau \tau^\prime}~W_{M}^{\tau \tau^\prime} +\tilde{R}_{\Sigma^{\prime \prime}}^{\tau \tau^\prime} ~W_{\Sigma^{\prime \prime}}^{\tau \tau^\prime}  +   \tilde{R}_{\Sigma^\prime}^{\tau \tau^\prime} ~ W_{\Sigma^\prime}^{\tau \tau^\prime} \right] \nonumber \\  
&&+ {q_\mathrm{eff}^{~2} \over m_N^2} ~\left[ \tilde{R}_{\Phi^{\prime \prime}}^{\tau \tau^\prime} ~ W_{\Phi^{\prime \prime}}^{\tau \tau^\prime}  
+   \tilde{R}_{\tilde{\Phi}^\prime}^{\tau \tau^\prime}~W_{\tilde{\Phi}^\prime}^{\tau \tau^\prime}  + \tilde{R}_{\Delta}^{\tau \tau^\prime}~ W_{\Delta}^{\tau \tau^\prime}  \right]\nonumber \\
&&-  {2 q_\mathrm{eff} \over m_N}~\left[ \tilde{R}_{ \Phi^{\prime \prime}M}^{\tau \tau^\prime} ~W_{ \Phi^{\prime \prime}M}^{\tau \tau^\prime}
 +  \tilde{R}_{\Delta \Sigma^\prime}^{\tau \tau^\prime} ~W_{\Delta \Sigma^\prime}^{\tau \tau^\prime} \right]  \Bigg\}~~~
\end{eqnarray}
where $M_T$ is the target mass, and  ${\phi}^{Z_\mathrm{eff}}_{1s}$ is a Coulomb factor described below.
This result agrees with the general rate formula deduced from symmetry considerations \cite{HRMR} and provides microscopic forms for the nuclear response functions $W_{\mathcal{O}}^{\tau\tau'}(q_\mathrm{eff})$, allowing them to be evaluated in the shell model. These response functions can be viewed as ``dials" that an experimentalist can tune by selecting appropriate targets.

The operator notation is standard in semi-leptonic weak interactions 
\cite{Liam2,TWD}, corresponding to charge $M_J$, longitudinal spin $\Sigma^{\prime \prime}_J$, transverse electric spin $\Sigma_J^\prime$, and transverse
magnetic velocity $\Delta_J$ operators.   The longitudinal $\Phi^{\prime \prime}_J$ and transverse electric $\tilde{\Phi}^\prime_J$ operators are generated from the spin-velocity current $\vec{v}_N \times \vec{\sigma}_N$.
CP and parity limit the contributing $J$ to even ($M_J$, $\Phi^{\prime \prime}_J$, $\tilde{\Phi}^\prime_J$) or odd ($\Sigma^{\prime \prime}_J$, $\Sigma_J^\prime$, $\Delta_J$) values.
The multipole operators are defined in \cite{HRMR}.  Taking $m_N \rightarrow \infty $ in Eq. (\ref{eq:rate}) yields the point-nucleus rate.

The $\tilde{R}_{\mathcal{O}}^{\tau\tau'}$ are our key result, as they define what can be learned from elastic $\mu \rightarrow e$ conversion:
\allowdisplaybreaks
\begin{eqnarray}
\label{eq:Rs}
 \tilde{R}_{M}^{\tau \tau^\prime} &=& \tilde{c}_1^\tau \tilde{c}_1^{\tau^{\prime} * } + \tilde{c}_{11}^\tau \tilde{c}_{11}^{\tau^\prime * }  \nonumber \\
 \tilde{R}_{\Phi^{\prime \prime}}^{\tau \tau^\prime}&=& \tilde{c}_3^\tau \tilde{c}_3^{\tau^\prime * } + \left( \tilde{c}_{12}^\tau- \tilde{c}_{15}^\tau\right) \left( \tilde{c}_{12}^{\tau^\prime *}-\tilde{c}_{15}^{\tau^\prime *} \right)  \nonumber \\
 \tilde{R}_{\Phi^{\prime \prime} M}^{\tau \tau^\prime} &=& \mathrm{Re} \left[ \tilde{c}_3^\tau \tilde{c}_1^{\tau^\prime *} -  \left( \tilde{c}_{12}^\tau - \tilde{c}_{15}^\tau \right) \tilde{c}_{11}^{\tau^\prime *}\right] \nonumber \\
  \tilde{R}_{\tilde{\Phi}^\prime}^{\tau \tau^\prime}&=& \tilde{c}_{12}^\tau \tilde{c}_{12}^{\tau^\prime * }+\tilde{c}_{13}^\tau \tilde{c}_{13}^{\tau^\prime *}   \nonumber \\
  \tilde{R}_{\Sigma^{\prime \prime}}^{\tau \tau^\prime} &=&  ( \tilde{c}_4^\tau- \tilde{c}_6^{\tau })  ( \tilde{c}_4^{\tau^\prime *}- \tilde{c}_6^{\tau^\prime *}) +\tilde{c}_{10}^\tau  \tilde{c}_{10}^{\tau^\prime *}   \nonumber \\
  \tilde{R}_{\Sigma^\prime}^{\tau \tau^\prime} &=&\tilde{c}_4^\tau \tilde{c}_4^{\tau^{\prime} * } + \tilde{c}_{9}^\tau \tilde{c}_{9}^{\tau^\prime * } \nonumber \\
    \tilde{R}_{\Delta}^{\tau \tau^\prime}&=&  \tilde{c}_{5}^\tau \tilde{c}_{5}^{\tau^\prime *}  + \tilde{c}_{8}^\tau \tilde{c}_{8}^{\tau^\prime *} \nonumber \\
 \tilde{R}_{\Delta \Sigma^\prime}^{\tau \tau^\prime}&=& \mathrm{Re} \left[ \tilde{c}_5^\tau \tilde{c}_4^{\tau^{\prime} * } + \tilde{c}_{8}^\tau \tilde{c}_{9}^{\tau^\prime * }   \right].
\end{eqnarray}
A program of $\mu \rightarrow e$ conversion experiments can in principle place eight constraints on the LECs.  Four CLFV ET operators,
 $\CO_2$, $\CO_7$, $\CO_{14}$, and $\CO_{16}$, are not probed due to parity and CP constraints.
 
\begin{figure}   
\centering
\includegraphics[scale=0.6]{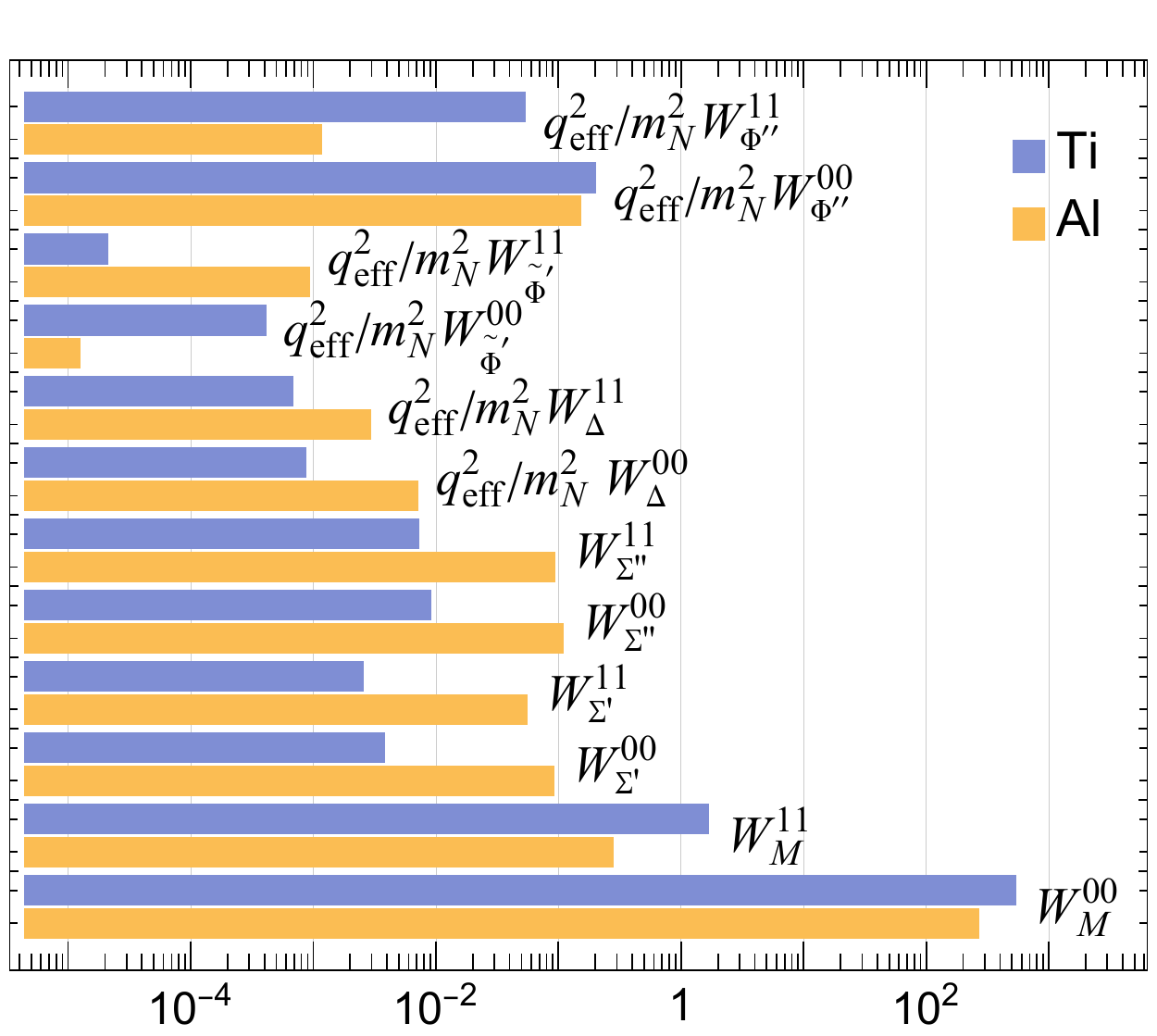}
\caption{Comparison of Al and Ti response functions, which govern sensitivities to the
CLFV bilinears defined in the text.}
\label{fig:Sens}
\end{figure}

Limits on the LECs can be obtained from experiment, using shell-model calculations of the nuclear response functions. As the CLFV LEC space has 32 degrees of freedom, we turn on each LEC separately
to assess sensitivity along each operator ``axis" in this space.  The LEC limits obtained from the existing Ti branching ratio bound \cite{TiLim} and from 
a projected $^{27}$Al limit of  $10^{-17}$ are shown in Table \ref{tab:LEClimits}.   Targets differ in sensitivity to the various operators, reflecting aspects of their ground-state structure.
We compare the strength of nuclear responses in Al and Ti in Fig. \ref{fig:Sens}. The next-generation target $^{27}$Al proves to be versatile in its range of sensitivities \cite{HRMR}. 

The tabulated  LEC limit can be converted into a CLFV mass scale through the relation
\be
 \Lambda_i^\tau \sim {\left(\sqrt{2}G_F\right)^{-1/2} \over {\sqrt{|\tilde{c}_i^\tau|}}}  \sim {246.2 \mathrm{~GeV} \over \sqrt{|\tilde{c}_i^\tau|}} 
 \ee
The LECs  $\tilde{c}_1^0$ and $\tilde{c}_{11}^0$, for which the nuclear response is fully coherent,
will be probed by Mu2e and COMET at scales $\gtrsim10^4$ TeV, while most of the remaining couplings are tested at levels $\sim 10^3$ TeV.
 
\begin{table}[!]
\label{table:limits}
\caption{LEC limits imposed by the indicated  $\mu \rightarrow e$ conversion branching ratios.
 E-$x \, \equiv \,10^{-x}.$}
\label{tab:LEClimits}
\begin{tabular}{|c||c|c||c|c|}
\hline
 & \multicolumn{4}{c|}{Target (Branching Ratio)} \\
 \hline
 \rule{0pt}{2.5ex}
 & \multicolumn{2}{c||}{Al ($10^{-17}$)} & \multicolumn{2}{c|}{Ti (6.1E-13)}\\
LEC& $\tau=0$ & $\tau=1$ & $\tau=0$ & $\tau=1$\\
\hline
$|\tilde{c}_1|,|\tilde{c}_{11}|$ &~4.0E-10~ & ~1.2E-8~ & ~7.4E-8~ & ~1.3E-6~ \\
$|\tilde{c}_3|,|\tilde{c}_{15}|$ & 1.6E-8 & 1.9E-7 & 3.8E-6 & 7.3E-6\\
$|\tilde{c}_4|$ & 1.4E-8 & 1.7E-8 & 1.5E-5 & 1.7E-5 \\
$|\tilde{c}_5|,|\tilde{c}_8|$ &7.8E-8 & 1.2E-7 & 5.8E-5 & 6.5E-5\\
$|\tilde{c}_6|,|\tilde{c}_{10}|$ & 2.0E-8 & 2.2E-8 & 1.8E-5 & 2.0E-5\\
$|\tilde{c}_9|$ & 2.1E-8 & 2.8E-8 & 2.8E-5 & 3.4E-5\\
$|\tilde{c}_{12}|$ & 1.6E-8 & 1.4E-7 & 3.8E-6  & 7.3E-6\\
$|\tilde{c}_{13}|$ & 1.8E-6 & 2.1E-7 & 8.4E-5  & 3.7E-4\\
\hline
\end{tabular}
\end{table} 

{\it Nuclear physics, enhancement, selection rules:} Nuclear response functions were evaluated using fully correlated shell-model wave functions constructed
from Slater determinants in a harmonic oscillator basis.  As the calculations are performed in complete spaces without truncation, we are able
to project out spurious center-of-mass motion, preserving Galilean invariance. Wave functions for  $^{27}$Al  and Ti were computed using BIGSTICK \cite{CWJ1,CWJ2} and the USDB $2s1d$  \cite{USDB} and KB3G $2p1f$ \cite{KB3G} interactions, respectively.  The Ti calculations were
summed over stable isotopes, weighted by their abundance.  The adopted oscillator parameters $b$ of 1.84 and 1.99 fm for Al and Ti, respectively,
are consistent with the r.m.s. charge radii of these nuclei.


The nuclear embedding alters the NRET:  First, the nucleus can enhance interactions due to coherence.  It is well known that the isoscalar monopole operator $M_0$  is coherently enhanced, and 
thus when it is present it dominates the rate.  In our NRET, a second monopole charge operator $\Phi^{\prime \prime}_0$ is generated from the spin-velocity 
current $\vec{\sigma}_N \times \vec{v}_N$, which is associated with tensor-mediated interactions such as
\be 
\bar{\chi}_e i \sigma^{\mu \nu} \gamma^5 \chi_\mu ~ \bar{N} i \sigma_{\mu \nu} \gamma^5 N .
\ee
Enhancement occurs in nuclei where one shell of a spin-orbit pair is filled and the other empty:  this is the case for Al  (filled $1d_{5/2}$, empty $1d_{3/2}$).  Although the pseudotensor interaction above generates a coupling to both nuclear charge and spin---so that $\vec{v}_N$-dependent contributions might be
expected to be a small correction---they in fact double the rate \cite{HRMR}.

Second, in elastic $\mu \rightarrow e$ conversion, nuclear selection rules can blind one to certain NRET operators.   For example, multipoles of the
axial charge $\vec{\sigma}_N \cdot \vec{v}_N$ violate either P or CP and thus cannot contribute to the ground-state nuclear process. Such operators are probed only in
inelastic $\mu \rightarrow e$ conversion.  

{\it Response functions and $\vec{v}_\mu$:}  The response functions in Eq. (\ref{eq:rate}) involve the convolution of a multipole nuclear density with
a gently varying muon wave function: in Al, the muon's Bohr radius $\sim$ 20 f.   
\begin{eqnarray}
W_{O}  &\equiv& {4 \pi \over 2j_N+1} \sum_J | \langle j_N  || \hat{O}_J^g(q) ||  j_N \rangle|^2 \nonumber \\
\hat{O}_J^g &=& \sum_{i=1}^A {1 \over \sqrt{4 \pi}} g(x_i) \hat{O}_{JM}(q x_i) 
\end{eqnarray}
A common practice, employed in Eq. (\ref{eq:rate}), replaces the muon wave function by an average value, 
\begin{eqnarray}
W_{O}  &\rightarrow &|\phi_{1s}^{Z_\mathrm{eff}}|^2  {4 \pi \over 2j_N+1} \sum_J | \langle j_N  || \hat{O}_J(q) ||  j_N \rangle|^2 \nonumber \\
\hat{O}_J &=& \sum_{i=1}^A  \hat{O}_{JM}(q x_i) 
\end{eqnarray}
The most frequently used procedure for performing this averaging is borrowed from the inclusive process of muon capture and is not optimal \cite{HRMR}. We define $\phi_{1s}^{Z_\mathrm{eff}}$ to reproduce the exact value of the monopole charge amplitude and
then use this value for all other response functions.  In Al, the average error induced in the rate by this approximation is 1.8\% \cite{HRMR}, much
less than typical nuclear structure errors in evaluating the response functions. While it is not strictly necessary, the use of $\phi_{1s}^{Z_\mathrm{eff}}$ greatly simplifies the nuclear physics, permitting all nuclear matrix elements to be expressed as polynomial functions of $y$.  

The NRET can be extended to include all effects linear in $\vec{v}_\mu$ \cite{HRMR}, yielding a small correction to the nuclear responses
\be
 W_O\rightarrow \frac{4\pi}{2j_N+1}\sum_J|\braket{j_N||\hat{O}^g_J+\hat{O}^{'f}_J||j_N}|^2
 \ee
where the new term is defined in analogy with $\hat{O}^g_J$ with $g(r)\rightarrow f(r)$. The form of the rate is unchanged: $\vec{v}_\mu$ only alters the nuclear form factors at the level $\sim 2\left|\braket{f}/\braket{g}\right|\sim 5\%$ in $^{27}$Al. In the literature, $\vec{v}_{\mu}$ has been included by truncating the multipole expansion at leading order, a procedure that, depending on the operator, can introduce errors that exceed the intended relativistic muon correction \cite{HRMR}.
Technical details of the work reported here, a discussion of the related CLFV processes $\mu\rightarrow e\gamma$ and $\mu \rightarrow 3e$, and a description of publicly-available Mathematica and Python scripts that can be used for EFT analysis may be found in \cite{HRMR}.  

\begin{acknowledgments}
The authors would like to thank Michael J. Ramsey-Musolf for extensive discussions. This work was supported in part by the US Department of Energy under grants DE-SC0004658, DE-SC0015376, and DE-AC02-05CH11231,
by the National Science Foundation under cooperative agreements 2020275 and 1630782, and by the Heising-Simons Foundation under award 00F1C7.
\end{acknowledgments}

\end{document}